\documentclass[fleqn,usenatbib]{mnras}

\usepackage{graphicx}   
\usepackage{amsmath}    
\usepackage{amssymb}    
\usepackage{txfonts}
\usepackage{natbib,longtable,times,multicol}
\usepackage{caption}
\usepackage[caption = false]{subfig}
\usepackage{verbatim}
\usepackage{makecell}
\usepackage{epstopdf}
\usepackage{bm}

\newcommand{\solarmass}{M_{\odot}}

\title[Environment and disc sizes]{On the influence of environment on star forming galaxies}

\author[Lizhi Xie et al.]{
Lizhi Xie,$^{1}$\thanks{E-mail: lzxie@oats.inaf.it} 
Gabriella De Lucia,$^{1}$
David J. Wilman,$^{2,3}$
Matteo Fossati,$^{2,3}$
Peter Erwin,$^{3}$
\newauthor
Leonel Guti\'errez,$^{4}$
Sandesh K. Kulkarni$^{3}$
\\
$^{1}$INAF - Astronomical Observatory of Trieste, via G.B. Tiepolo 11, 
I-34143 Trieste, Italy\\
$^{2}$Universit{\"a}ts-Sternwarte M{\"u}nchen, Scheinerstrasse 1, D-81679 M{\"u}nchen, Germany\\
$^{3}$Max-Planck-Institut f{\"u}r Extraterrestriche Physik, Giessenbachstrasse, D-85748 Garching, Germany\\
$^{4}$Instituto de Astronom\'ia, Universidad Nacional Aut\'onoma de M\'exico, Apdo. Postal 877, Ensenada, 22800 Baja California, Mexico.
}

\date{Accepted XXX. Received YYY; in original form ZZZ}

\pubyear{2017}

\begin{document}

\label{firstpage}
\pagerange{\pageref{firstpage}--\pageref{lastpage}}
\maketitle

\begin{abstract}
  We use our state-of-the-art semi analytic model for GAlaxy Evolution and
  Assembly (GAEA), and observational measurements of nearby galaxies to study
  the influence of the environment on the gas content and gaseous/stellar disc
  sizes of star forming galaxies.  We analyse the origin of differences between
  physical properties of satellites and those of their central counterparts,
  identified by matching the $V_{\rm max}$ of their host haloes at the
  accretion time of the satellites. Our model reproduces nicely the differences
  between centrals and satellites measured for the HI mass, size of the star
  forming region, and stellar radii. In contrast, our model predicts larger
  differences with respect to data for the molecular gas mass and star formation 
  rate. By analysing the progenitors of central and satellite model galaxies,
  we find that differences in the gas content arise after accretion, and can be
  entirely ascribed to the instantaneous stripping of the hot gas reservoir. 
  The suppression of cold gas replenishment via cooling and star formation lead 
  to a reduction of the cold gas and of its density. Therefore, more molecular gas 
  is lost than lower density HI gas, and model satellites have less molecular gas 
  and lower star formation rates than observed satellites. We argue that these 
  disagreements could be largely resolved with the inclusion of a proper treatment 
  for ram-pressure stripping of cold gas and a more gradual stripping of the hot gas 
  reservoir. A more sophisticated treatment of angular momentum exchanges, accounting
  for the multi-phase nature of the gaseous disc is also required.

\end{abstract}




\section{Introduction}
\label{sec:introduction}

It has long been known that galaxy properties correlate with their environment:
galaxies in dense regions of the Universe are redder, more passive and more
concentrated than those in regions with `average' density
\citep[e.g.][]{dressler1980,balogh1999,poggianti1999,lewis2002,gomez2003,
  kauffmann2004,bamford2009,peng2012}.  Complementary trends are found if one
focuses on the abundance of gas in galaxies
\citep{bothun1980,chamaraux1980,giovanelli1985,
  solanes2001,boselli2002,boselli2006,koopmann2004,kenney2004,chung2009,odeken2016,jaffe2016}. HI
deficiencies (i.e. the lack of HI with respect to isolated galaxies of similar 
morphological size and optical size) are typically ascribed to
environmental effects. This hypothesis, however, is difficult to test 
as it would require, in principle, the identification of the progenitors of
galaxies observed today, at a time when they were residing in similar
environments.

Theoretically, there are a number of physical processes that can effectively
reduce the cold gas content of galaxies in dense environments: (i)
`strangulation', i.e. the removal of the hot diffuse gas reservoir associated
with galaxies falling into denser structures \citep{larson1980}; (ii)
`ram-pressure stripping' of cold gas suffered by galaxies travelling at large
velocities through the diffuse intra-cluster medium \citep{gunn1972}; (iii)
`tidal stripping' due to the gravitational interaction with the parent halo or
with other galaxies \citep{merritt1983}.  (iv) `galaxy harassment', i.e. the
effect associated with repeated high-velocity encounters, which is believed to
play a role in the formation of dwarf ellipticals or the destruction of low
surface brightness galaxies in clusters \citep{moore1996}.  The efficiency of
these processes at different scales has been studied extensively using detailed
numerical simulations
\citep[e.g.][]{tonnesen2009,tecce2010,guo2011,steinhauser2016,emerick2016,stevens2017}. Their
relative importance in driving the observed environmental trends remains,
however, debated.

Recent studies have combined observational measurements with simulated accretion 
histories to constrain the timescales for the suppression of star formation in 
satellite galaxies (related to the timescale necessary to significantly deplete 
their cold gas reservoir). These are rather long ($\sim$ 3-8 Gyr in the local 
Universe), with a dependence on both galaxy stellar mass and redshift
\citep[e.g.][]{delucia2012,wetzel2013,hirschmann2014,fossati2017}. Although
these results should be interpreted in a `probabilistic' sense (not all
galaxies will shut off simultaneously, and the scatter in quenching timescale
is likely correlated with the orbital distribution of infalling galaxies),
these long timescales are difficult to reproduce in theoretical models of
galaxy formation \citep[e.g.][]{hirschmann2014,bahe2015,luo2016,brown2017}.

The ratio between the sizes of the star forming and stellar discs ($R_{\rm
SFR}/R_{\star}$) of satellite galaxies can provide important information
about environmental processes: ram-pressure is expected to remove first
low-density gas at large distance from the galaxy centre. The stellar disc
would be unaffected, at least until tidal stripping of stars becomes
effective. Therefore, `weak' ram-pressure stripping should cause a
decrease of the HI size but would likely not affect significantly the
molecular and stellar disks, leading to no significant evolution of the
$R_{\rm SFR}/R_{\star}$ size ratio. `Strong' ram-pressure stripping can
affect the low-density HI and also the denser molecular gas in the disk,
causing a decrease of the $R_{\rm SFR}/R_{\star}$ size ratio, up until the
point at which the galaxy is completely quenched (i.e. no longer forms stars,
even in the centre). In case of strangulation, the density of cold gas is
expected to decrease at all radii, due to star formation and stellar
feedback, so that the ratio between the sizes of the gaseous and stellar disc
should remain approximately constant initially. The decreasing gas density 
leads to a lower molecular-to-atomic gas ratio and therefore a lower star 
formation efficiency, that becomes negligible at large radii. The shrinking 
star forming region therefore leads to an increase of the size ratio between 
the HI and the stellar or molecular disc.

Statistical studies focusing on the size-mass relation have so far 
mainly focused on stellar disc sizes. These have indicated that late-type 
galaxies in dense environments are slightly more concentrated (have smaller 
sizes) than those in the field \citep{weinmann2009,kuchner2017,spindler2017}. 
Multi-wavelength surveys have allowed us to gather important information on 
how the gas content of individual galaxies is affected by the environment. 
Based on studies of galaxies in nearby clusters, \citet{cortese2012} and 
\citet{fossati2013} showed that HI-deficient galaxies have star forming discs 
smaller than stellar discs, and that the size ratio decreases with HI-deficiency.  
A large fraction of HI-deficient late type galaxies are also depleted in molecular 
hydrogen, i.e. the star forming reservoir \citep{boselli2002,fumagalli2009}. 
The depletion of HI is, however, more efficient than that of star forming gas 
\citep{fabello2012,catinella2013}. Because of stripping galaxies in a dense 
environment can have truncated HI density profiles \citep{cayatte1990,cayatte1994}, 
 molecular profiles \citep{fumagalli2009,boselli2014c} dust profiles
\citep{cortese2010}, and H$_{\alpha}$ profiles 
\citep{kenney2004,koopmann2006,fossati2013,schaefer2017}.  In some cases, a tail of
HI and ionised gas is observed \citep{gavazzi1995,chung2009,jachym2017,bellhouse2017}, 
which can be interpreted as a consequence of ram-pressure \citep{tonnesen2010}.

Semi-analytic models of galaxy formation have been crucial to improve our
understanding of the correlation between galaxy properties and their evolving
environment.  \citet{xie2015} found that the size-mass relation of early-type
central galaxies correlates tightly with the formation time of their host
haloes: $R_{\star}\approx H(z(t_f))^{-2/3}$.  On the basis of this result, we
argued that the evolution of the stellar size, at fixed stellar mass, can be 
explained by differences in the halo assembly histories. These are expected 
to be large when comparing central and satellite galaxies: host haloes of 
satellite galaxies have likely formed earlier than those hosting central 
galaxies of the same stellar mass, and have suffered significant stripping 
after being accreted. In this work, we will combine observational estimates 
with state-of-the-art semi-analytic models to explore the origin of the 
observed size differences between centrals and satellite star forming galaxies today.

This paper is organized as follows: Section~\ref{sec:modelanddata}
introduces the semi-analytic model used in this study and the
observational samples considered.  In Section~\ref{sec:gas}, we
compare observed and predicted integrated properties of central and
satellite galaxies at $z=0$, and explain the differences between
centrals and satellites by studying their evolution histories.  In
Section~\ref{sec:size}, we review the prescriptions adopted to model
disc sizes, and compare observational measurements and model
predictions for the sizes of central and satellite galaxies. In
Section~\ref{sec:discussion}, we discuss our results, that are then
summarized in Section~\ref{sec:conclusion}.

\section{The galaxy formation model and Observational data}
\label{sec:modelanddata}
In this section, we introduce the model and the observations that we
use in this paper, and discuss how we select model galaxies to be
compared with data.

\subsection{The galaxy formation model}
\label{subsec:sam}

In this work, we take advantage of the latest version of our GAlaxy Evolution
and Assembly (GAEA) model. GAEA features a sophisticated chemical enrichment
scheme \citep{delucia2014} that accounts for the non instantaneous recycling
of gas, metals and energy from massive stars and different types of supernovae,
and a new stellar feedback scheme based partly on results from hydrodynamical
simulations \citep{hirschmann2016}. The version of GAEA we use in this work
also includes an explicit treatment for the partition of cold gas in molecular
(H$_2$) and atomic hydrogen (HI), and H$_2$-based star formation laws
\citep{xie2017}. Specifically, we use here the prescriptions based on the
empirical relation by \citet[BR06 in \citealt{xie2017}]{blitz2006}, as this
model provides a better agreement with different observational data,
including the $M_{\rm HI}-M{\star}$, $M_{\rm H_2}$-$M_{\star}$ scaling
relations and the HI and H$_2$ mass functions measured in the local Universe.

For the present analysis, we apply our model to the Millennium II simulation
\citep[][MSII]{boylankolchin2009}. This corresponds to a box of $100\,{\rm
  Mpc}\,{\rm h}^{-1}$, simulated employing a particle mass of
$6.89\times10^6\solarmass\,{\rm h}^{-1}$. The simulation assumes a WMAP1
cosmology, with $\Omega_{m}=0.25$, $\Omega_{b}=0.045$, $\Omega_{\Lambda}=0.75$,
$h=0.73$, and $\sigma_8=0.9$.  More recent measurements from e.g. PLANCK
\citep{planck2015} and WMAP9 \citep{bennett2013} provide slightly different
cosmological parameters and, in particular, a larger value for $\Omega_{m}$ and
a lower one for $\sigma_8$.  As shown in previous work, however, these
differences are expected to have little influence on model predictions, once
the parameters are retuned to reproduce a given set of observables in the local
Universe \citep{wang2008,guo2013}. As discussed in \citet{xie2017} and in our
previous work, our model applied to the MSII can resolve well galaxies with
stellar mass $M_{\star}>10^8 M_{\odot}$.

For the following discussion, it is worth noting that our model
assumes that the hot gas associated with galaxies infalling on larger systems
(i.e. becoming satellites) is instantaneously stripped. Satellite galaxies
can continue forming stars until their reservoir of cold gas is
exhausted. Finally, we do not include a modelling for the stripping of cold gas
due to ram-pressure \citep{gunn1972}.

From the model, we randomly select 6000 `star forming' satellite
galaxies. These are selected by fitting the predicted SFR-stellar mass
relation for satellites (we find: $\log SFR_{\rm MS, m} = 0.9\log 
M_{\star}-8.8$), and then considering only galaxies with $\log
{\rm SFR} > \log {\rm SFR_{MS,m}}-1$.  For each of these satellite
galaxies, we also select a central star forming galaxy (using the same
`star forming' definition) among those with $V_{\rm max}$ comparable
to the halo $V_{\rm max}$ of the satellite galaxy at the time it was
accreted (i.e. at the last time it was a central
galaxy). Specifically, we require $-0.1 < \delta \log V_{\rm max,acc}
< 0.5$
\footnote{An asymmetric range is used because of the non-uniform
  distribution of halo masses.}, where $\delta \log V_{\rm max,acc}$
is the difference between the halo $V_{\rm max}$ of the central and
satellite galaxies at the satellite's accretion time.  Since in our
model we keep $V_{\rm max}$ fixed to the value the satellite galaxy
had at the last time the galaxy was central, this allows us to match
centrals and satellites that were in similar `environments'
before environmental effects start playing a role.

The selected sample is representative of star forming satellites. These
  have been accreted between $z\sim 1$ and $z\sim 0.02$ and, at present, they
  reside in haloes with mass ranging between $10^{10}$ and $3\times 10^{14}
  M_{\odot}$ with median $\sim 10^{13} M_{\odot}$.  Their central counter-parts are typically hosted by lower mass
  haloes (ranging between $10^{10}$ and $3\times 10^{14} M_{\odot}$, with median 
  halo mass $\sim 2\times 10^{11} M_{\odot}$.)
depending on the stellar mass. \footnote{ The satellite population selected from 
the model follows a similar distribution of halo masses as satellite galaxies 
from the HAGGIS and the xGASS. Satellites in the HRS are 
located in the Virgo cluster. For central galaxies, model galaxies are hosted
on average by lower mass haloes than those in observations.}

\subsection{Observational data}
\label{subsec:data}

In this work, we use data from the {\it Herschel} Reference Survey
  (HRS hereafter - \citealt{boselli2010}), the extended GALEX Arecibo
SDSS Survey (xGASS hereafter - \citealt{catinella2018}) and from the
H$\alpha$~Galaxy Groups Imaging Survey (HAGGIS hereafter -
\citealt{Kulkarni2014}).

HRS is a volume-limited, $K$-band-selected sample of 322 local galaxies
(15<dist<25 Mpc), including fairly isolated objects and galaxies within the Virgo 
cluster. In this paper, we focus on late-type star forming galaxies and exclude $22$ 
galaxies that are classified as elliptical galaxies. The stellar mass is calculated 
from $i$-band luminosity and $g-i$ colour \citep{cortese2012,zibetti2009}. The star 
formation rates (SFRs) used in this paper are obtained by averaging four different 
estimates by \citet{boselli2015} and correcting for dust attenuation. The HRS also 
includes atomic hydrogen masses from ALFALFA \citep{giovanelli2005,haynes2011} and 
\citet{springob2005} and molecular hydrogen mass \citep{boselli2014a}. The latter  
are obtained from CO(1-0) luminosity using a constant conversion factor 
$X_{CO} = 2/3\times 10^{20} {\rm cm^{-2}/(K\, (km\,s^{-1}))}$. The effective radii of 
the star forming and stellar components of each galaxy corresponds to the half-light radii of 
the H$_{\alpha}$ emission and in the $r$-band, respectively.  We exclude $68$ galaxies with 
no information of sizes or SFR. This selection leaves us with $233$ galaxies, including $54$ in
the core of the Virgo cluster, $59$ falling onto the cluster and $120$
non-Virgo galaxies \citep{gavazzi1999}. We consider galaxies in the cluster
core and those infalling as `satellites', and those non-Virgo galaxies as
`centrals' (this might include some non-Virgo satellite galaxies in the
sample of central galaxies).

HAGGIS is a narrow band H$\alpha$ imaging survey that includes $390$ galaxies 
located in over $100$ galaxy groups with halo mass ranging between $10^{12}$ 
and $10^{14}\solarmass$. These have been selected from the Sloan Digital Sky 
Survey (SDSS) group catalogue by \citet{yang2007}.  Galaxies are classified 
as centrals or satellites based on the original Yang et al. classification. 
The $i$-band luminosities are also based on SDSS data. The stellar masses 
of HAGGIS galaxies are estimated from the $g-i$ colour and $i$-band luminosity 
as for the HRS galaxies. The SFR is estimated from the extinction-corrected 
H$_\alpha$ luminosity. HAGGIS does not provide information for gas masses,
and includes both star forming and quiescent galaxies.  For our
analysis, we exclude galaxies with no detection in H$_{\alpha}$ and no assignment 
of halo mass. AGN galaxies are also excluded from HAGGIS for the inaccurate estimating 
on their SFRs and SFR radii.

xGASS is a gas fraction-limited census of 1179 galaxies in the local 
Universe ranging between $10^9<M_{\star}$ and $10^{11.5} M_{\odot}$. The
HI-detections are from GASS \citep{catinella2013}, GASS-low \citep{catinella2018}, 
and ALFALFA \citep{haynes2011}. We use only the non-confused HI-detected galaxies. 
Each galaxy is classified as a central or a satellite based on the SDSS group 
catalogue by \citet{yang2007}. The stellar mass is taken from the SDSS MPA/JHU 
catalogue.  The SFR is estimated from the ultra-violet 
\citep[UV, from GALEX -][]{martin2005, morrissey2007} and mid-infrared (MIR) 
luminosity \citep[{\it from Wide-field Infrared Survey Explorer -}][]{wright2010}, 
or SED fitting \citep{wang2011}.  xCOLDGASS \citep{saintonge2017} provides CO 
detections for 290 galaxies and upper limits for 122 galaxies. We convert the 
CO(1-0) luminosity to H$_2$ mass using a constant conversion factor as done for 
HRS galaxies. xGASS provides half mass radii based on r-band imaging, but no 
information for the SFR radii. Therefore, we do not include xGASS in our analysis of 
galaxy sizes. 

All properties of observed galaxies are corrected to a Chabrier IMF and 
WMAP 1yr cosmology as used in our galaxy formation model.

\begin{figure*}
\includegraphics[width=1.\textwidth]{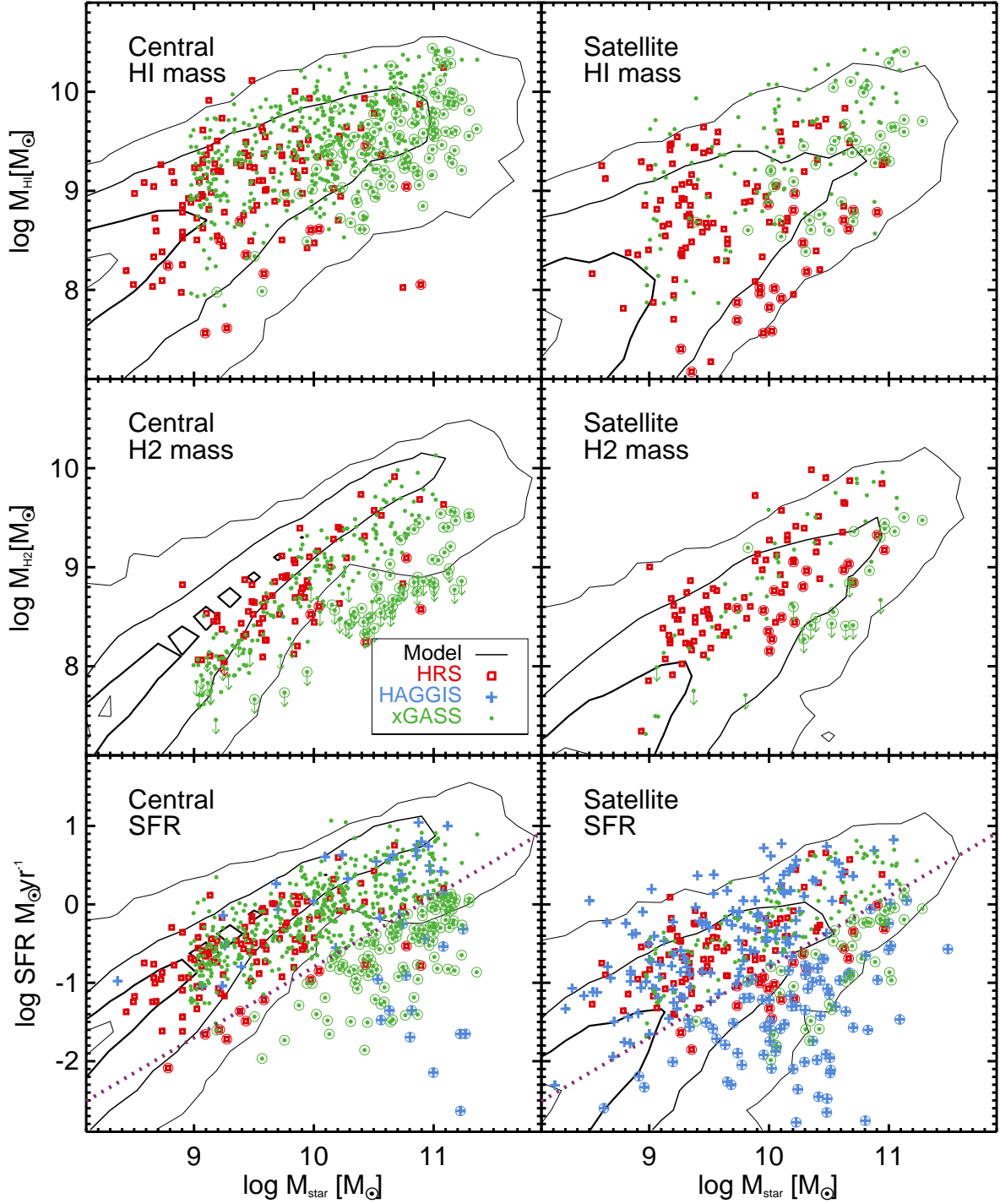}
  \caption{Distributions of HI mass (top), H$_2$ mass (middle) and SFR
    (bottom) for central (left) and satellite (right) galaxies as a
    function of stellar mass. The black contours show the number
    density of all central/satellite galaxies in the model, with thin
    to thick lines indicating number densities corresponding to $1$,
    $10$, $50$ per cent of the peak density. The red squares, green
    dots and blue crosses represent galaxies from HRS, xGASS and
    HAGGIS, respectively. Symbols with open circles represent passive 
    galaxies that are more than 1 dex below the main sequence. 
    Symbols with downward arrows are upper limits. The  purple dotted line
    in the bottom panels shows the SFR cut adopted to select star
    forming galaxies from the data and the model. }
 \label{fig:distribution}
\end{figure*}

In Fig.~\ref{fig:distribution}, we plot the distribution of
integrated properties for all galaxies from the HRS (red squares),
xGASS (green dots), HAGGIS (blue crosses), and all model galaxies 
(black contours) at $z=0$.  Most of the HRS galaxies have 
stellar masses below $\sim 10^{10}\,{\rm M}_{\sun}$, while xGASS and
HAGGIS include relatively more massive galaxies.  Measurements from
the different surveys are consistent with each other, and our model
reproduces the observed distributions relatively well, both for
centrals and for satellite galaxies.

The top and middle panels show the HI mass and H$_2$ mass as a function 
of galaxy stellar mass.  Model galaxies follow the same distribution
as galaxies from HRS and xGASS on the HI mass - stellar mass plane. 
The middle panels show, instead, that the model slightly over-predicts 
the H$_2$ masses of central galaxies and slightly under-predicts the 
H$_2$ masses of satellites.  The bottom panels shows the relation between 
SFR and stellar mass. HRS and xGASS include more galaxies with high SFR 
than HAGGIS. Model galaxies occupy the same region of data for both central 
and satellite galaxies when compared to galaxies from HRS, but predict a 
narrower distribution at large stellar masses with respect to galaxies from 
xGASS and HAGGIS. We select star forming galaxies from the data using the 
same SFR cut as the one used for model galaxies (shown as a purple dotted 
line in bottom panels of Fig.~\ref{fig:distribution}). {We have tested lower SFR cuts and also a different (empirical) MS relation \citep{speagle2014} to select star forming galaxies. We find that our results are not affected by these different choices qualitatively.}

\section{Integrated properties of central and satellite galaxies}
\label{sec:gas}

In this section, we analyse the differences between integrated 
properties of central and satellite galaxies, and compare our model 
predictions with observational data.

\subsection{Differences at present time}
\label{subsec:diffnow}

\begin{figure*}
\includegraphics[width=1.\textwidth]{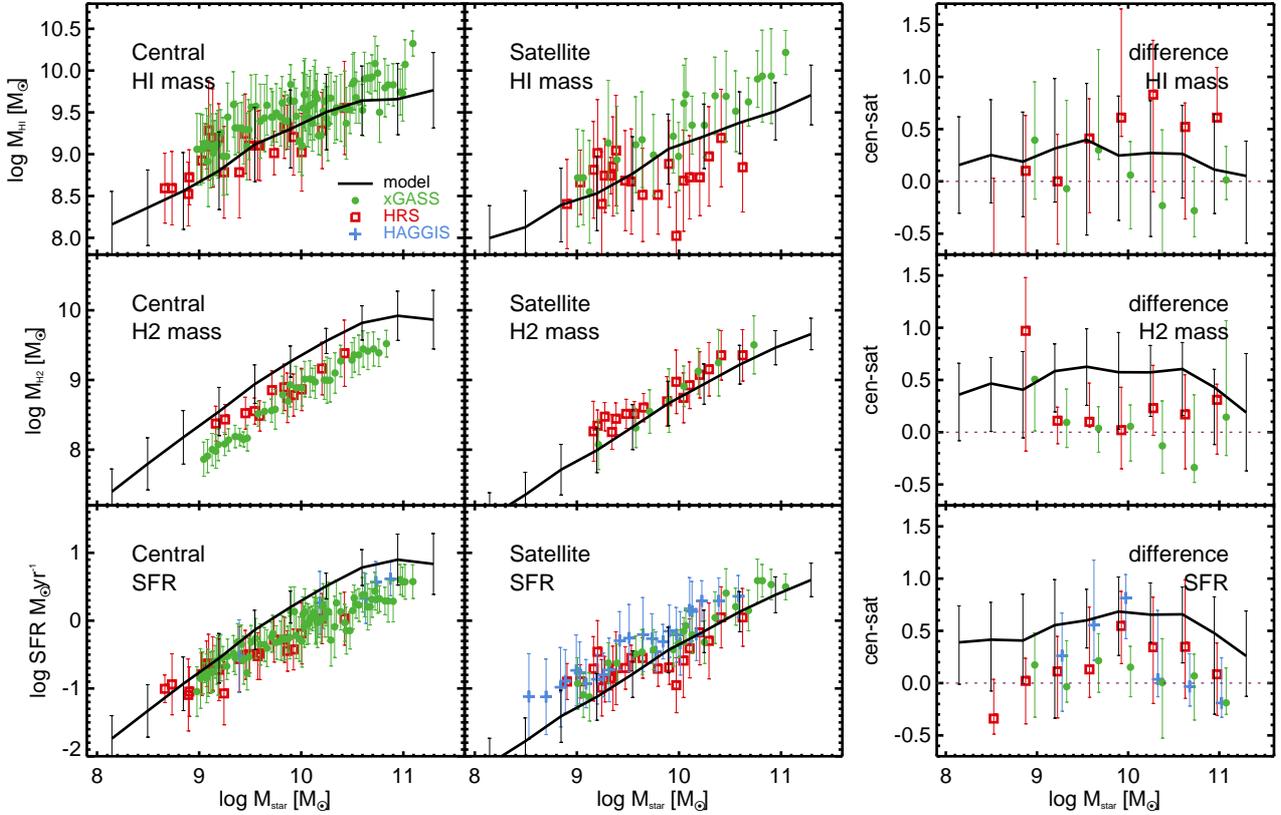}
\caption{The top, middle, and bottom rows show the median HI mass, H$_2$ mass,
  and SFR as a function of galaxy stellar mass for centrals (left column) and
  satellites (middle column). The right column shows the differences between the
  integrated properties of central and satellite galaxies. Red squares, blue
  crosses, and green dots show the data from HRS, HAGGIS and xGASS,
  respectively. The solid black lines show our model predictions. In the left
  and middle column, black lines and error bars correspond to median values and
  standard deviations, while coloured symbols are running medians with a window
  width of $9$ galaxies.  In the right column, black lines and coloured 
  symbols show median differences and error bars the 9th and 98th percentiles  
  (more details in the text). The purple line is shown as a reference.}
   \label{fig:compare}
\end{figure*}

 Fig.~\ref{fig:compare} shows the median scaling relations of central (left 
column) and satellite (middle column) star forming galaxies. The black lines 
show the median and standard deviation of the distributions predicted by our 
model.  The coloured symbols are running medians computed from the observational 
samples. Specifically, we have sorted the galaxies in the samples by their stellar 
mass, and computed the median properties and standard deviations in bins containing 
$9$ galaxies \footnote{Changing the number of galaxies in the calculation of the 
running median, or in the estimation of the difference between centrals and satellites 
described later in the text, does not affect our results qualitatively.}. Our model 
reproduces reasonably well the distribution observed for HI masses of both central 
and satellite galaxies. For the molecular gas content and the SFR, the agreement 
is less good. Specifically, the model over-predicts by $0.3$ dex the H$_2$ mass 
of central galaxies, and under-estimates by $\sim 0.1$ dex that of satellites galaxies. 
In addition, the model predicts a steeper SFR-M$_{\star}$ relation than observed, 
which is a common problem for semi-analytic models (e.g. see discussions in 
\citet{xie2017,cora2018}). 
In order to account for inconsistencies between model predictions and observational measurements for central galaxies, and focus on environmental effects, we concentrate below on the differences between centrals and satellites.

The right column shows the differences between central and satellite galaxies.
To estimate the uncertainty of these differences, we randomly select $9$
galaxies in a given stellar mass bin from the central and satellite samples,
and compute the difference between the median value of each sub-samples. 
Results change significantly by repeating the procedure. In order to give a 
conservative estimate of the scatter, we repeat the selection $50$ times and 
show as error bars the 2nd and 98th percentiles of the distributions of the 
estimated differences.  We find that, both for our model galaxies and for the 
observational samples, star forming satellite galaxies have less gas and lower 
SFR than central galaxies of similar stellar mass. Specifically, we find that 
the model central-satellite difference is $\sim 0.2$ dex for the HI mass and 
more than a factor of two larger ($\sim 0.5$ dex) for the H$_2$ mass and the 
SFR. The central-satellite differences for the observed galaxies show a larger 
variation. The xGASS and HRS are roughly consistent with each other and give a 
central-satellite difference for the HI mass that is comparable to that predicted. 
The differences measured for the H$_2$ mass and SFR are instead smaller 
($< 0.2$ dex) than those predicted by our model. The consistency between model predictions and observational estimates is worse if a lower SFR selection cut is adopted. This is due to a more rapid 
depletion of the molecular gas content (and therefore a more rapid suppression 
of the SFR) in the model with respect to the data. As discussed in the 
introduction, this is a problem shared by most existing theoretical models of 
galaxy formation \citep[see e.g.][]{hirschmann2014,brown2017}.  We will further 
discuss possible solutions to this problem in Sec.~\ref{sec:discussion}.

\subsection{Evolution of the difference between central and satellite galaxy}

We then take advantage of our model results to understand how and when the
differences between centrals and satellites measured today are established. 
To this aim, we compare the properties of galaxies at fixed halo $V_{\rm 
max,acc}$, i.e. at fixed values of $V_{\rm max}$ at the accretion time of the 
satellite galaxies. We assume that $V_{\rm max,acc}$ is representative of 
the `environment' of the galaxy at accretion time, so that centrals and 
satellites in haloes with similar $V_{\rm max,acc}$ are assumed to have evolved 
in similar environments before accretion.

Since we have selected centrals by matching their $V_{\rm max,acc}$ 
with that of the satellites at the time of accretion, the stellar mass of 
satellite galaxies will be, in general, different from that of their central 
counterparts. Specifically, we find that the difference in stellar mass ranges 
from $0.6$ to $-0.4$ dex for galaxies in haloes from low $V_{\rm max.acc}$ to 
high $V_{\rm max.acc}$. To remove the dominant trend with galaxy stellar mass, 
we compare the difference between `normalised' properties: Fig.~\ref{fig:compare} 
shows that the scaling relations for HI mass, H$_2$ mass, and SFR for model 
galaxies are well fit by $M_{\rm HI} \propto M_{\star}^{0.6}$, $M_{\rm H2} 
\propto M_{\star}^{0.9}$, and $SFR \propto M_{\star}^{0.9}$. We assume the 
slopes of these relations do not vary with cosmic time and normalize the HI 
mass, H$_2$ mass, and SFR by $M_{\star}^{0.6}$, $M_{\star}^{0.9}$, and 
$M_{\star}^{0.9}$, respectively.

\begin{figure*}
\includegraphics[width=1.\textwidth]{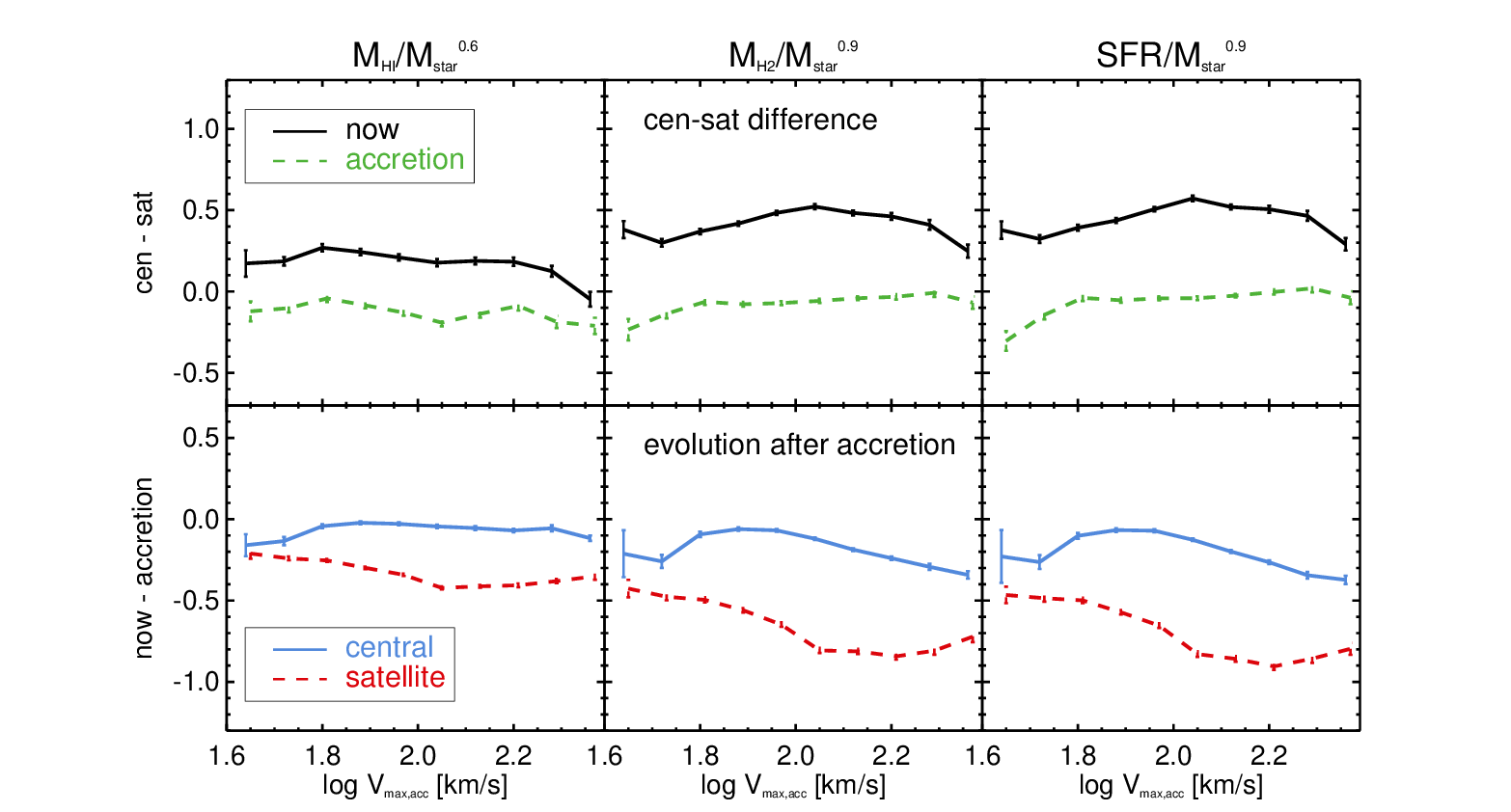}
\caption{Top panels show the differences between the normalized HI mass 
  (left column), H$_2$ mass (middle column), and SFR (right column) of 
  centrals and satellites predicted at present (solid black lines) and 
  at the time of accretion (green dashed lines). The bottom quantities
  show instead the difference between properties predicted at present 
  and at the time of accretion. Solid (blue) and dashed (red) lines are 
  used for central and satellite galaxies, respectively.
   Lines correspond to the median of the distributions while error bars
  show the errors on the mean.}
\label{fig:prog_sat}
\end{figure*}

The top left panel of Fig.~\ref{fig:prog_sat} shows the differences
between the normalised HI mass of central and satellite galaxies at accretion
time (green dashed) and now (solid black). At present time, central galaxies 
have $0.2$ dex more HI than their satellite counterparts (this is similar to 
the difference shown in Fig.~\ref{fig:compare} when comparing central and 
satellite galaxies at fixed stellar mass).  The difference at accretion time 
ranges from $-0.1$ to $-0.3$ dex.  Therefore, the difference between normalised 
HI mass of central and satellite galaxies has increased \textbf{by} about $0.3-0.4$ dex 
since accretion.

The evolution of the central-satellite difference is explained by the
bottom left panel, showing the difference between the normalized HI mass for
centrals (solid blue lines) and satellites (red dashed lines) predicted at
present and at the time of accretion. For central galaxies, the normalized HI
mass remains unchanged since accretion. For satellite galaxies, the normalized
HI mass decreases by up to $0.4$ dex, with a mild dependence on the $V_{\rm
  max,acc}$. The reason for this differential evolution is that satellite
galaxies lose HI due to environmental effects (stripping of the hot gas
reservoir in our case). The trend with $V_{\rm max,acc}$ can be explained by
the fact that satellites accreted in more massive haloes have been accreted
on average earlier than satellites in lower mass haloes, and therefore these 
was more time to deplete their cold gas reservoir

The middle and right columns show results for the normalised H$_2$ mass and
SFR. At present time, the difference of both H$_2$ normalized mass and
normalized SFR at fixed $V_{\rm, max, acc}$ is about $0.3-0.4$ dex. At
accretion time, the difference varies between $-0.2$ dex and zero. After 
accretion, the difference has increased by $\sim 0.5$ dex. The bottom panels 
show the differences between normalised H$_2$ mass/SFR of galaxies at present 
and accretion time. Central galaxies (blue solid lines) lose $\sim 0.2$ dex of
their normalized H$_2$ mass and SFR since accretion, which corresponds to the
redshift evolution of H$_2$ mass - stellar mass relation and SFR - stellar mass
relation.  For satellite galaxies (red dashed lines), the decrease of both
H$_2$ normalized mass and normalized SFR is more pronounced, particularly for
haloes that are more massive at the time of accretion: the decrease is up to
$0.8$ dex in the $V_{\rm max,acc}$ range we have considered. 

To summarize, the model and observational data predict consistent trends for
satellite galaxies with small difference in the amplitude. The model predicts that central and
satellite galaxies which evolved in similar environment before accretion have
slightly different integrated gas properties. Specifically, central galaxies
have slightly less HI mass, H$_2$ mass, and lower SFR than satellite galaxies
on average. As we have discussed, the difference between integrated properties
of centrals and satellites observed at $z=0$ can be ascribed to environmental
effects.

\section{Difference in sizes of gaseous and stellar discs}
\label{sec:size}

In this section, we focus on the sizes of gaseous and stellar discs. We
discuss how disc sizes and angular momenta of central and satellite star
forming galaxies evolve in the framework of our semi-analytic model, and
how the predicted differences between central and satellite galaxies
compare to observational measurements. We then trace the evolution of our
model galaxies to determine when such differences arise and why.

\subsection{Disc sizes and angular momentum evolution in the model}
\label{sec:discsizes}

Galaxies sizes are modelled by tracing the angular momentum of the gaseous and
stellar components, as described in \citep{guo2011}. We assume that gaseous and
stellar discs are rotationally supported, in equilibrium, and described by
exponential density profiles. Under these assumptions, their scale lengths are
given by \citep{mo1998}: 
\begin{equation}
r_{\rm gas,d} = \frac{J_{\rm gas}/M_{\rm gas}}{2V_{\rm max}}, \\\
r_{\rm \star,d} = \frac{J_{\rm \star}/M_{\rm\star}}{2V_{\rm max}},
\label{eqn:size}
\end{equation}
where $J_{\rm gas}$, and $J_{\star}$ are the angular momenta of the gaseous and
stellar disc, respectively. $M_{\rm gas}$, and $M_{\star}$ are the total mass
of gas and stars in the disc. $V_{\rm max}$ is the maximum circular velocity
and is computed from the actual distribution of dark matter particles in the
haloes associated with model galaxies. In the case of satellite galaxies, this
quantity corresponds to that of the parent halo at the last time the galaxy was
central.

The stellar and gaseous discs of galaxies gain mass and angular momentum
through various physical processes. We assume that the specific angular
momentum of the hot gas (i.e. the diffuse baryons that are shock heated during
accretion onto the dark matter halo) is identical to that of the host dark
matter halo. Recent work based on hydro-dynamical simulations
\citep[e.g.][]{stevens2017,danovich2015} has shown that the gas accreted
through the `rapid cooling regime' (effective at high redshift and in small
haloes) can carry an angular momentum from two to four times larger than that
of the parent dark matter halo. In \citet{zoldan2018}, we show that this
does not influence significantly the size of late-type model galaxies in the
local Universe.

The gaseous disc then gains or loses angular momentum and mass through: (i) gas
cooling, that transfers gas from the hot halo to the cold gas disc; (ii) star
formation, that turns a fraction of the cold gas into stars; (iii) recycling,
that turns a fraction of the stars back into cold gas; (iv) supernovae
feedback, that heats a fraction of the cold gas up to the virial temperature of
the halo, and ejects a fraction of the heated gas out into the inter-galactic
medium; (v) mergers, that bring gas from satellite galaxies onto centrals
\footnote{Our model also includes satellite-satellite mergers, i.e. mergers
  between `orphan' galaxies and satellite galaxies still associated with a
  distinct dark matter substructure. These are, however, rare.} and can convert
(depending on the mass-ratio) part of the remnant gas into stars. The same
physical processes, with the exclusion of gas cooling and supernovae feedback,
affect the angular momentum of the stellar disc. In addition, this component is
affected by disc instabilities. This process moves some stars from the inner
disc to a non-rotational bulge, which decreases the mass of the stellar disc
while conserving its angular momentum. Therefore, in our models, disc
instability events increase the specific angular momentum and size of the
stellar disc.

The variations of angular momentum for the gaseous ($\Delta \bm{J}_{\rm gas}$)
and stellar ($\Delta \bm{J}_{\star}$) disc at each code time-step, can be
written as:
\begin{equation}
\Delta \bm{J}_{\rm gas}=\bm{J}_{\rm cooling} - \bm{J}_{\rm SF} 
+ \bm{J}_{\rm  recycling} + \bm{J}_{\rm merger,gas} - \bm{J}_{\rm SNfb}, 
\label{eqn:Jgas}
\end{equation}
and
\begin{equation}
\Delta \bm{J}_{\star}= \bm{J}_{\rm SF} - \bm{J}_{\rm recycling} 
+ \bm{J}_{\rm merger,\star} +  \bm{J}_{\rm inst}
\label{eqn:Jstar}
\end{equation}

In the following, we define the size of the gaseous disk as that enclosing 
half of the total gas mass and the size of the stellar component as the radius 
enclosing half of the stars in the bulge+disk. The stars and the gas in the disc 
are assumed to follow an exponential distribution with scale-length given by 
Eq.~\ref{eqn:size}, while those in the bulge are assumed to be distributed 
according to a \citet{jaffe1983} profile. 

We partition the gaseous disk in an HI and an H2 component, and form stars 
only from the latter. As explained in detail in Xie et al. (2017), the partition 
is performed in 21 annuli, which allows us to compute the extent of the star 
forming region. In the following, we define the SF size as the radius enclosing 
half of the total SFR for each model galaxy. In our model, the distribution of 
the newly formed stars follows that of the entire gaseous disk (HI+H2) at the 
time of the star formation, but the amount of stars formed depends on the estimated 
molecular-to-atomic gas ratio. Eliminating this inconsistency requires a more
realistic treatment of angular momentum exchanges, that accounts for
the multi-phase nature of the cold gaseous disc. This goes beyond
the aims of this work, and we plan to address this problem in future work.

\subsection{Contribution to disc growth from different physical processes}
\label{subsec:discgrow}
\begin{figure*}
\includegraphics[width=1.0\textwidth]{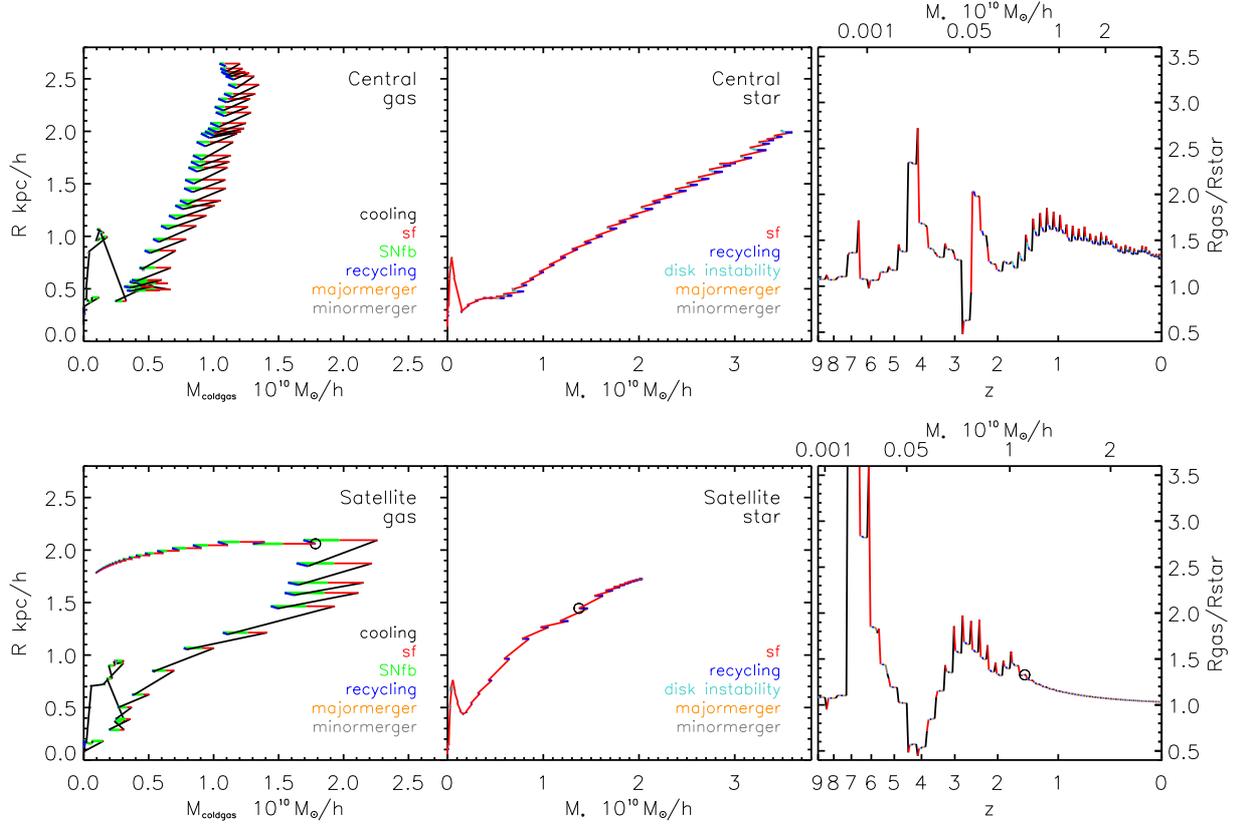}
\caption{The growth of two example galaxies in the size-mass
  plane. The top panels correspond to a disc dominated central galaxy,
  while the bottom panels correspond to a satellite galaxy. The left
  (middle) panels show the evolution of the scale length of the
  gaseous (stellar) disc as a function of the gas (stellar) mass.  The
  insets show a zoom in on the late time evolution. The grey circle in
  the bottom panels marks the accretion time for the satellite galaxy
  (i.e. the time corresponding to the last time the galaxy is
  central).  The right panels show how the ratio between the scale
  radius of the gaseous and stellar disc evolves as a function of
  redshift. In all panels, coloured lines show variations of the
  quantities shown corresponding to different physical processes (as
  labelled), and occurring at each code time-step. Gray thin curves in
  the right panels connect size ratios measured at subsequent
  snapshots.}
\label{fig:size_growth}
\end{figure*}

In Fig.~\ref{fig:size_growth}, we show the growth of the gaseous and stellar
disc radii for two example galaxies (a central galaxy in the top panels, and a
satellite galaxy in the bottom ones). These have been randomly selected among
model galaxies with stellar bulge-to-total mass ratio smaller than $0.3$, and
stellar mass $M_{\star}\sim 10^{10}\solarmass$. The left panels show the
evolution of the gaseous disc scale length as a function of the gas mass. Each
segment corresponds to a change in disc scale length and/or gas mass, and is
colour-coded according to the physical processes that has driven that
variation. The middle panels show the corresponding evolution in the size-mass
plane for the stellar component of the disc\footnote{As explained in the
  previous section, the stellar size includes the bulge. So, strictly speaking,
  we are looking here at the stellar component of the entire galaxy, and not of
  the disc. However, these galaxies are selected to be disc-dominated so that
  the stellar size of the disc is generally very close to that of the
  galaxy.}. The figure shows that the gas disc mainly evolves (both in mass and
in size) through gas cooling. The gas disc size can increase or decrease due to
cooling depending on the instantaneous value of the dark matter halo
spin. Stellar feedback decreases the cold gas mass (moving a fraction of it
into the hot gas reservoir), but it does not affect the size of the disc
because the specific angular momentum of the gas is conserved. Therefore,
time-steps during which the evolution is driven by supernovae feedback are
shown as horizontal green lines. Recycling does not affect significantly the
mass of the disc, but it can significantly modify its size adding back into the
inter-stellar medium gas with lower specific angular momentum from previous
stellar populations. This is particularly notable at early times, so that
blue segments (showing time intervals during which the evolution is driven by
recycling) at early times are almost vertical. At later times, the variation in
mass due to recycling becomes more significant than the variation it causes in
size. The bottom left panel shows that, before accretion onto a more massive
system (this is marked by a black circle), the size evolution of satellite
galaxies follows similar trends as for central galaxies. After accretion, the
gas disc loses mass because of star formation and stellar feedback, and gains
mass through gas recycling. Since there is no additional gas cooling, and only
relatively small fractions of stars are formed after accretion, the specific
angular momenta of the gas and stellar discs (and therefore their sizes) are
not significantly modified after accretion.

The middle panels of Fig.~\ref{fig:size_growth} show the evolution of the
stellar disc size as a function of the stellar disc mass. The evolution of the
stellar disc size generally follows that of the corresponding gaseous disc,
with the evolution being driven primarily by star formation. What happens,
typically, is that the gaseous disc grows first due to cooling. Star formation
then transfers the angular momentum of the gas to the stellar component,
driving the growth of the stellar disc. This is displayed nicely in the right
panels, that show how the ratio between the scale length of the gaseous disc
and that of the stellar disc varies as a function of redshift. The size ratio
oscillates significantly, with gas cooling generally leading to an increase of
the gaseous disc size (and therefore an increase in the size ratio plotted),
and star formation increasing the size of the stellar disc (therefore
decreasing the size ratio). For the satellite galaxy examined, the size of the
stellar disc keeps growing (although very little) after accretion, and the
final size ratio is slightly larger than unity.

\begin{figure*}
\includegraphics[width=0.8\textwidth]{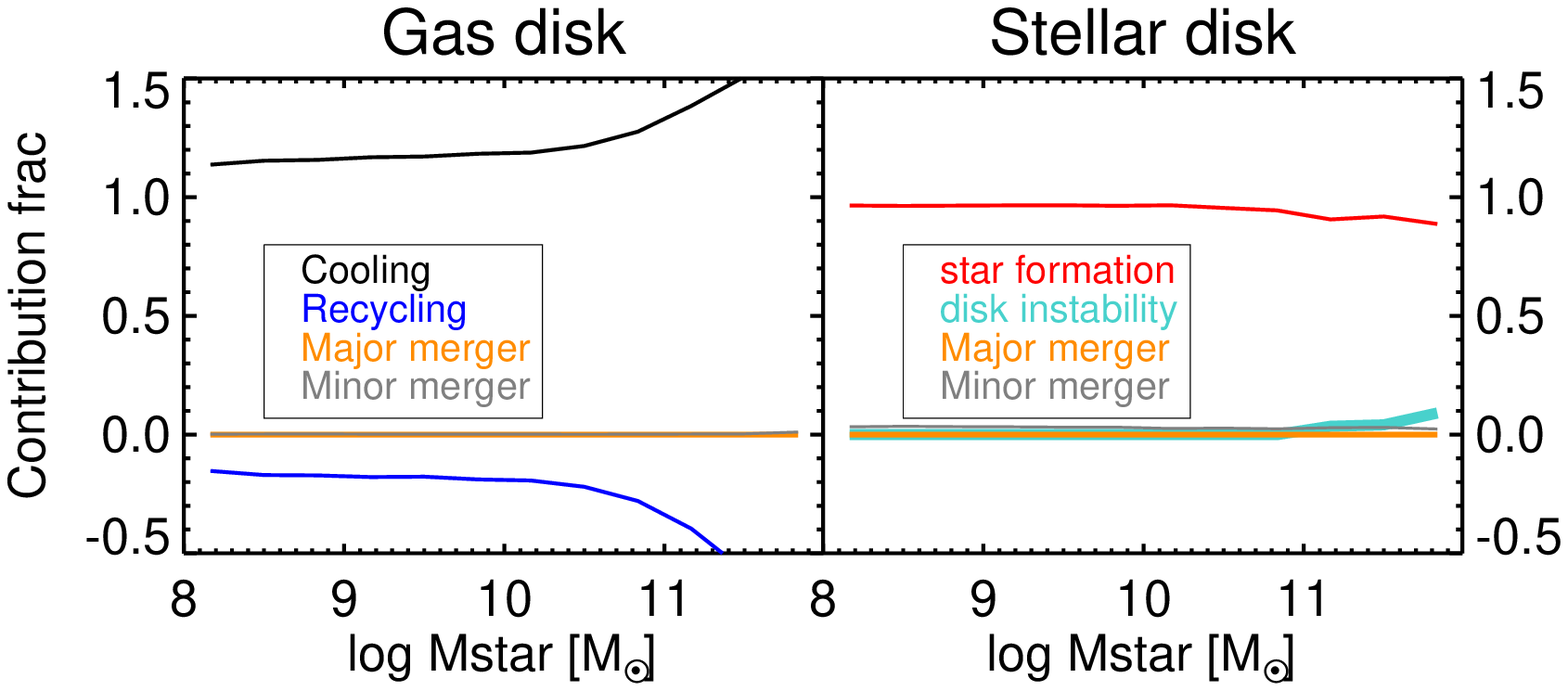}
\includegraphics[width=0.8\textwidth]{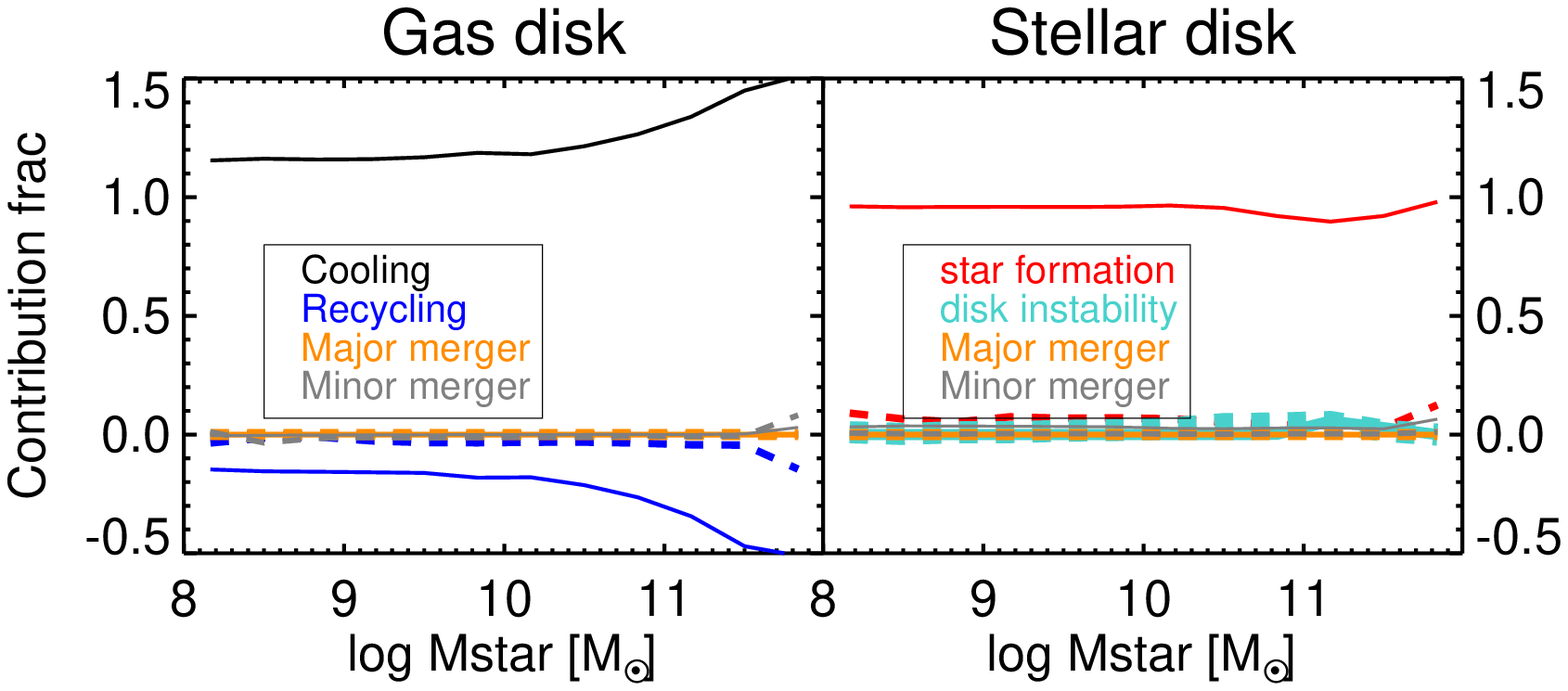}
\caption{Median fractional contribution of different physical processes in
  determining the final sizes of gas discs and stellar discs, as a function of
  the galaxy stellar mass. Top and bottom panels show results for central and
  satellite galaxies, respectively.  In the bottom panels, dashed lines
  correspond to the contributions by different physical processes after
  accretion. Only disc dominated galaxies have been included in the analysis
  ($B/T<0.3$).}
\label{fig:contribution_spiral}
\end{figure*}

Fig.~\ref{fig:contribution_spiral} summarizes the fractional contribution of
various physical processes to the growth of the gaseous (left column) and
stellar (right column) disc sizes. Top and bottom panels correspond to central
and satellite galaxies, respectively. The quantities plotted have been computed
averaging results for late-type ($B/T<0.3$) galaxies.  Let us focus first on
the top panels (central galaxies): the figure confirms that, on average,
cooling drives the growth of the gaseous disc size, with a trend for an
increasing contribution with increasing stellar mass. The contribution from
major mergers is negligible, which is not surprising given these galaxies have
been selected to be disc dominated. The contribution from minor mergers is also
very small, but somewhat more important for galaxies more massive than $\sim
10^{11}\,M_{\odot}$. Recycling on average gives a negative contribution to gas
disc growth, i.e. it tends to decrease the size of the gaseous disc because it
typically restitutes gas of lower specific angular momentum. The stellar disc
of central galaxies grows $\sim 90$ percent of its size through star
formation. This fraction is constant over the stellar mass range considered.
Minor mergers contribute for less than ten per cent of the stellar disc growth
size, for all galaxy stellar masses considered. Finally, disc instability
contributes very little for low-mass galaxies and is somewhat more important
(the contribution is never higher than 10 per cent) for more massive galaxies.

The bottom panels of Fig.~\ref{fig:contribution_spiral} show the corresponding
fractional contributions from different physical processes for satellite
galaxies. In this case, dashed lines show the contribution to size growth by
different physical processes after accretion ($z\sim 0.45$ on average).  The
results are consistent with contributions for central galaxies, indicating that
there is no obvious difference in size evolution between central and satellite
galaxies in our model. After accretion, both gaseous and stellar sizes grow
very little. In particular, stellar sizes grow due to star formation while the
gaseous disc sizes tend to decrease slightly due to gas recycling (and minor
mergers). 

\subsection{Differences of radii between central and satellite galaxies}
\label{subsec:diffsize}

\begin{figure*}
\includegraphics[width=1.\textwidth]{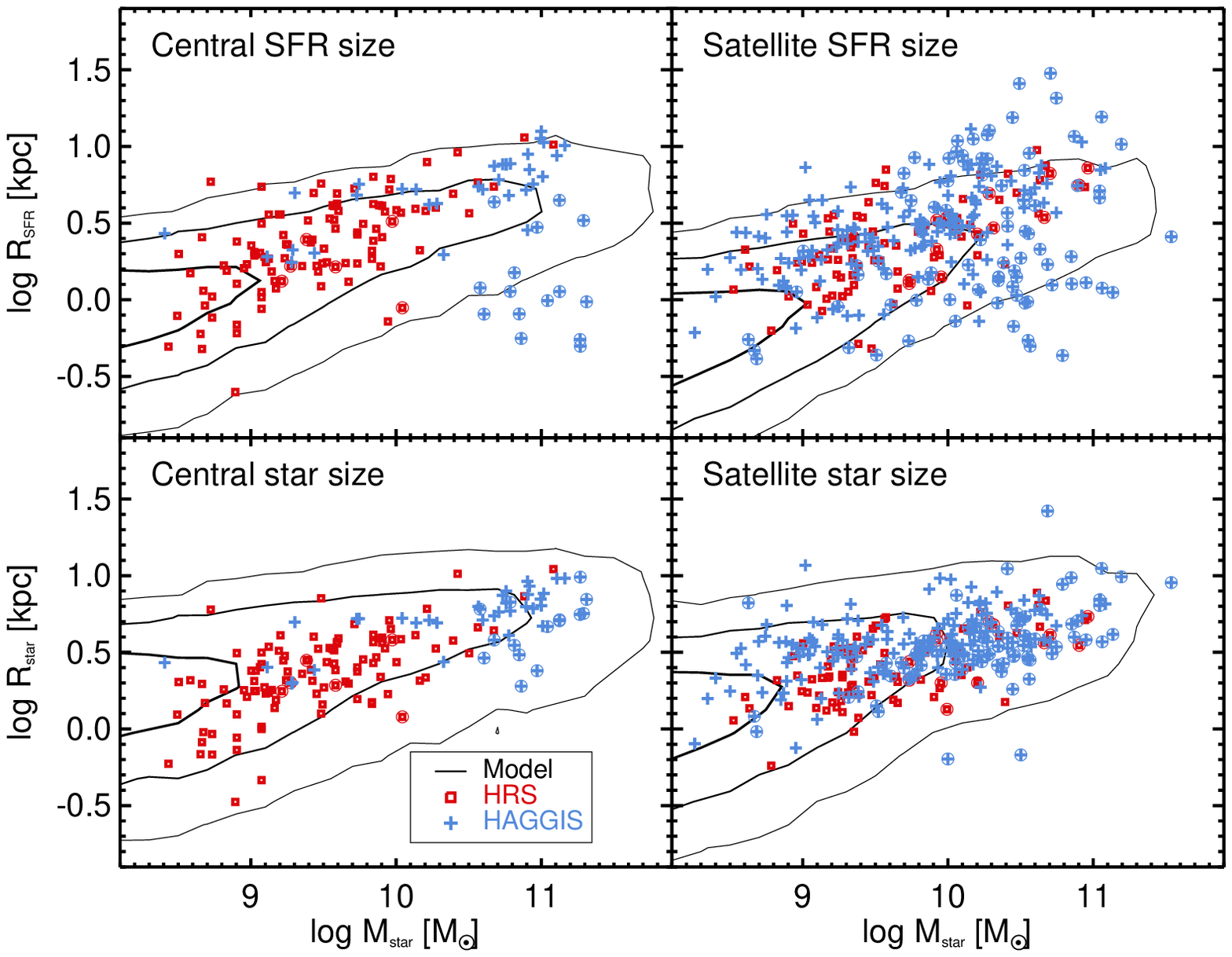}
\caption{Distributions of the size of the star forming region (top),
  stellar size (middle) and gas size (bottom) for central (left) and
  satellite (right) galaxies as a function of stellar mass. Symbols and contours 
  have the same meaning as in Fig.~\ref{fig:distribution}.
}
\label{fig:all_size}
\end{figure*}

\begin{figure*}
\includegraphics[width=1.\textwidth]{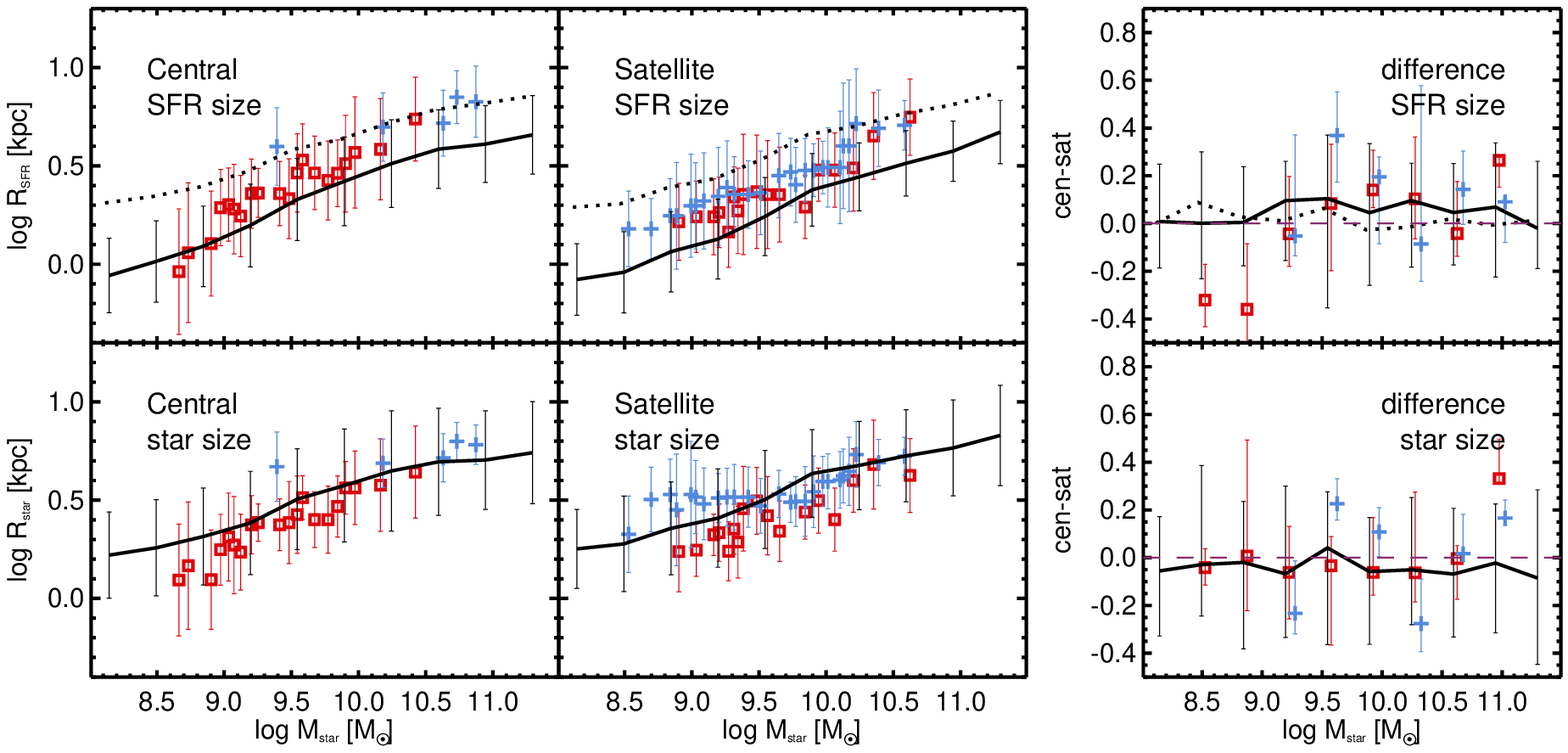}
\caption{The top and bottom panels show the median size of the star forming
  region and of the stellar disk as a function of galaxy stellar mass. Left and
  middle columns are for centrals and satellites, while the right column shows
  the difference between centrals and satellites. Symbols and lines have the same meaning as in
  Fig.~\ref{fig:compare}. The solid and dotted lines in the top panels
  correspond to the size of the model star forming region and of the model gaseous disk, respectively.}
\label{fig:compare_size}
\end{figure*}

We compare the stellar and SF radii of model galaxies with observational 
measurements in Fig.~\ref{fig:all_size}. Top and bottom rows show the distribution 
of SF and stellar radii respectively, as a function of the galaxy stellar mass. 
Left and right columns show results for central and satellite galaxies. Red squares 
and blue crosses represent individual galaxies from HRS and HAGGIS (we have included 
quiescent galaxies, marked with open circles, in this case). These two data 
sets exhibit consistent distributions for the SF and stellar sizes, with HAGGIS 
including a larger fraction of galaxies with low SFR at the massive end, which leads 
to a larger scatter in the distributions. The black contours show the distribution 
of all model galaxies. These occupy the same region of observed galaxies and exhibit 
a similar scatter.

We then select star forming galaxies from the model and the observational
samples, and compare their median radii at fixed stellar mass in
Fig.~\ref{fig:compare_size}. In the top panels, the median SF sizes of observed
galaxies lie between median values predicted for the SF radii and the gaseous
disc radii of model galaxies.  Median stellar radii of model galaxies are in
reasonable agreement with observational measurements. In the right column,
we plot the differences between central and satellite galaxies against stellar
mass. These differences have been calculated following same procedure described
for Fig.~\ref{fig:compare}. We find that model central galaxies have $\sim 0.1$ dex
larger SF radii than model satellite galaxies at fixed stellar mass. A similar
difference is found for galaxies in HRS and HAGGIS, although the scatter is
very large. We also plot the differences for the model gaseous disc sizes as dotted
lines in the top panels, and find that these are only slightly smaller than the
difference we predict for SF sizes.  Both in the model and in the data, central
and satellite galaxies have similar stellar radii at fixed stellar mass (bottom
right panel). 

As done for the integrated properties, we then analyse the evolution of central
and satellite galaxies, and compare the central-satellite differences at the
accretion time and now. In order to remove trends with galaxy stellar mass, we
normalise the sizes by $M_{\star}^{0.2}$. Indeed, Fig.~\ref{fig:compare_size}
shows that the relation between SF, gas or stellar radii and stellar mass is
well fitted by a power law with index $0.2$.

\begin{figure*}
\includegraphics[width=1.\textwidth]{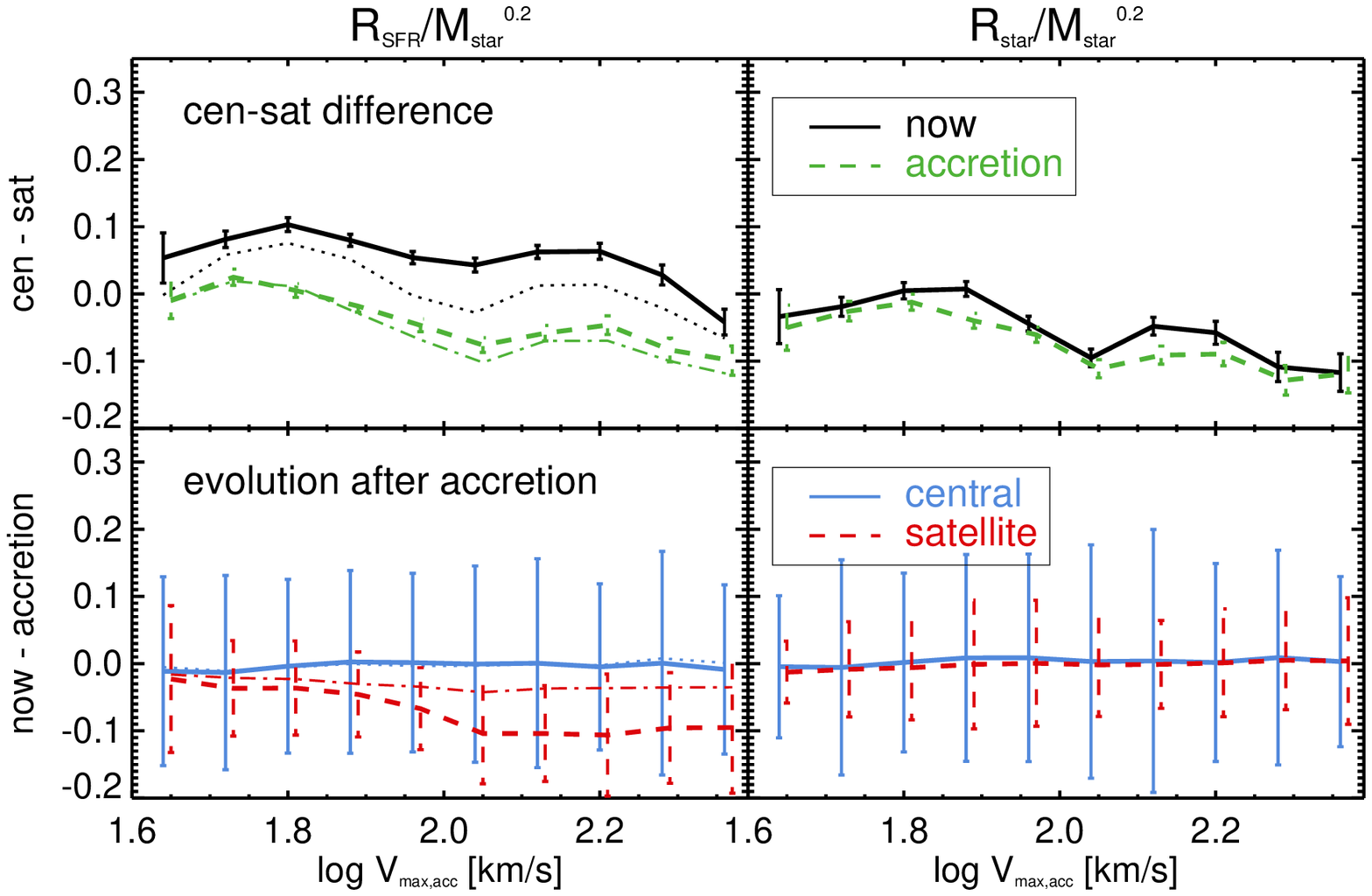}
\caption{Top panels show the differences between the normalized SF (left) and
  stellar (right) radii of central and satellite galaxies at present (black
  solid) and at accretion time (green dashed). In the left panel, black dotted
  and green dot-dashed lines show the differences for the gaseous disc radii.
  Bottom panels show the differences between the properties predicted at present
  and at the time of accretion. Solid blue and dashed red lines correspond to
  central and satellite galaxies, respectively. Blue dotted and red dot-dashed
  lines show results for gaseous disc radii of central and satellite galaxies.}
\label{fig:prog_size}
\end{figure*}

Top left panel of Fig.~\ref{fig:prog_size} shows the differences between the
normalized SF radii of satellite galaxies and those of their central counterparts.  
At present time, the SF radii of central galaxies are about $\sim 0.1$ dex larger
than those of their satellite counterparts. Differences are smaller for the gaseous 
disc radii. At the time of accretion, central galaxies have smaller SF radii than
their satellite counterparts, with a difference ranging between -0.1 and 0.03 dex.
Comparable differences are found for the gaseous disc radii. The difference between 
the SF radii of central and satellite galaxies has increased by $0.1$ dex since 
accretion time. For the gaseous disc radii (green dot-dashed line and black dotted 
line in the top-left panel) the difference between centrals and satellites has 
increased by a smaller amount. The bottom left panel shows the evolution of the 
normalized SF radii and of the gaseous disc radii for central and satellite galaxies. 
The evolution of the SF radii of central galaxies corresponds to the redshift 
evolution of the $R_{\rm SF}$ - stellar mass relation predicted by the model. 
Since accretion, the SF radii and the gaseous disc radii of central galaxies have 
increased only slightly. For satellite galaxies, the gaseous disc size remains 
stable, while the SF radii decreases by $\sim 0.1$ dex. As explained above, the 
evolution of the SF radius follows, in our model, tat of the entire gaseous disk. 
Once a galaxy is accreted, gas cooling (the main driver for the evolution of the 
gaseous disc, as discussed in Sec.~\ref{sec:discsizes}) is suppressed so that the 
gaseous disk size remains stable. The SF radii of satellite galaxies, however,
decreases due to star formation that slowly uses up the residual cold gas
reservoir. The gas density decreases, leading to a decrease of the H$_2$-to-HI
ratio at all radii, and eventually to a decrease of the star forming region
size. 

The right column shows the corresponding results for the stellar radii. The top
right panel shows that the difference between stellar radii of central and
satellite galaxies is very small. This is due to the fact that (i) the
difference at accretion is very small and (ii) the difference barely changes 
after accretion.  As we will discuss in more detail in
Sec.~\ref{subsec:hisize}, point (i) can be explained by the similar assembly
history of the host haloes of central and satellite galaxies before
accretion. Point (ii) can instead be explained by the lack of significant
evolution for the stellar radii of both centrals and satellites after accretion,
as shown in the bottom right panel.

Our results show that our model can reproduce the observed SF radii difference
between central and satellite galaxies considering only strangulation. The
central-satellite difference is partly due to the effect of
strangulation on satellite galaxies (i.e. it prevents the gaseous disk size from
growing further because cooling is suppressed), and partly a natural consequence
of the modelling we have adopted for the partition of cold gas in its atomic
and molecular gas components.

\section{Discussion}
\label{sec:discussion}

\subsection{Model limitations and possible improvements}

In Sec.~\ref{sec:gas}, we have shown that our model satellites deplete
  their H$_2$ reservoir more efficiently than their HI content. This results in
  a too rapid suppression of their star formation activity with respect to
  expectations based on observational data. We expect that ram-pressure
  stripping of cold gas, that is not accounted for in our current model
  version, and a non-instantaneous stripping of the hot gas reservoir, could
  bring our model predictions into better agreement with data. Below, we
  elaborate on this in more detail.

In the model, a satellite galaxy suffering strangulation loses gas (via star 
formation) with no modification of the gaseous disc size. The gas density 
decreases at all radii, which decreases the molecular fraction. So it is 
possible to find model satellites with large HI discs but no significant 
ongoing star formation. Observational data suggest that galaxies residing in 
denser environments (that can be identified as satellite galaxies) tend to be 
HI-deficient \citep[][and references therein]{boselli2006}, and are typically more depleted of
their HI content than molecular gas \citep{fumagalli2009,boselli2014c}. This is expected
because atomic hydrogen is typically more extended than H$_2$, and should be
more easily stripped from galaxies travelling at high speed through the diffuse
intra-cluster medium. So the expectation (and this seems supported by data) is
that the outer gas disc edge is truncated, while the central regions are
unaffected and still characterized by a high molecular ratio, and therefore
active star formation. In our model galaxies, the molecular fraction depends on
gas density so that the HI disc is indeed more extended than the molecular
disc. By including an explicit treatment for ram-pressure stripping 
of cold gas, and lowering the efficiency of strangulation, model galaxies 
should deplete their HI reservoir earlier (and more efficiently) than their 
molecular gas content. In this scenario satellite galaxies would also have 
longer quenching time-scales, because star formation can be sustained for longer 
times by the existing molecular reservoir, and by replenishment with new gas 
cooling from the hot gas reservoir. We plan to test this scenario in
  future work.

\subsection{Correlation between HI mass and gas size}
\label{subsec:hisize}

We find that central galaxies have smaller HI masses and gas radii than their 
satellite counterparts at accretion time (see Fig.~\ref{fig:prog_sat} and 
Fig.\ref{fig:prog_size}). The differences are due mainly to early-accreted 
galaxies. Our model satellites have been accreted between  $0<z_{\rm acc}<1$. We 
find that satellite galaxies that are accreted before $z\sim 0.5$ have larger 
gas disc radii, and larger HI mass than their central counterparts at the 
accretion time. The reason is the later formation time of haloes hosting the 
progenitors of these satellite galaxies with respect to those corresponding 
to their central counterparts.

Our model predicts a tight correlation between gas disc radii and HI mass for 
central galaxies, since both properties are related to the assembly history 
of their host haloes. One quantity that characterizes the halo assembly history 
is the `formation time' $f_{halo}$, typically defined as the time when the halo achieves 
half of its final mass  - see detailed discussion in 
\citealt{xie2015,zoldan2018}).

The HI mass depends on the halo assembly time, because the HI fraction 
depends on both the gaseous radius and the total amount of gas available. 
The latter is tightly correlated to the formation time: galaxies in 
early-formed haloes tend to have lower gas fraction compared to those in 
late-formed haloes (see \citealt{zoldan2018} for more detailed discussions). 
In addition, galaxies in early-formed haloes also have smaller sizes than 
those in late-formed haloes. In our model, the molecular-to-atomic gas ratio 
is proportional to the surface gas density:
\begin{displaymath}
  \frac{M_{\rm H2}}{M_{\rm HI}} \propto \Sigma^{2\alpha}_{\rm gas} \propto
  \left(\frac{M_{\rm gas}}{R^2_{\rm gas}}\right)^{2\alpha}
\end{displaymath}
with $\alpha =0.92$ (see \citealt{xie2017}). Therefore, galaxies in haloes
assembled at early redshifts have higher molecular ratios, and lower HI masses,
than those in haloes assembled later.

\begin{figure}
\includegraphics[width=0.5\textwidth]{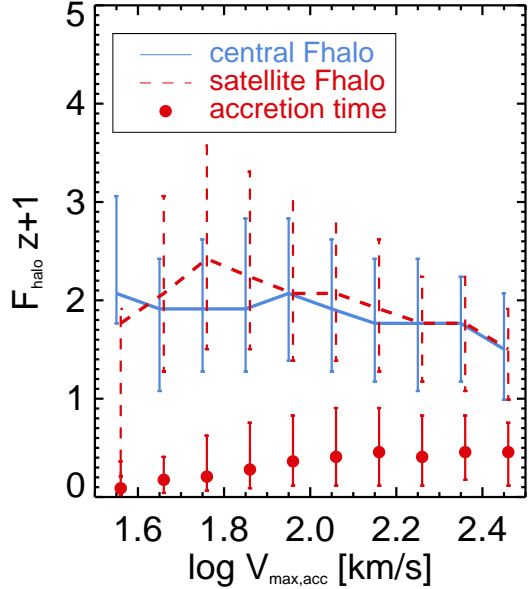}
\caption{Median formation time of haloes hosting our model galaxies. Blue solid
    lines and red dashed lines correspond to central and satellite galaxies,
    respectively. Error bars are the $16$, $84$ percent scatters.
    Red points show the median accretion times for haloes hosting
    satellite galaxies. }
\label{fig:fhalo}
\end{figure}

It is therefore worth noting that our results might depends on the 
selection. 
Most of our selected model central/satellite pairs have 
similar assembly histories. 
Fig.~\ref{fig:fhalo} shows the redshifts corresponding to 
the halo formation times as a function of halo $V_{\rm max,acc}$. 
We define the $f_{halo}$ as the time when the 
halo achieves half of its mass at accretion time.
The figure shows that the selected centrals and satellites indeed sit in haloes 
that are formed at comparable times. This leads to comparable SF/stellar radii
at accretion time. Therefore the differences of HI mass and sizes between 
central and satellite galaxies arise after accretion, and can be entirely 
ascribed to environmental effects. 

When using observational data, it is impossible to know if the central and
satellite galaxies were in similar environment at earlier epochs so that the
biases we have just described might become important.

\section{Conclusions}
\label{sec:conclusion}

We use the state-of-art semi-analytic model GAEA, together with observational
measurements from the HRS, xGASS and HAGGIS surveys to study the gas content
and SF/stellar disc sizes of star forming galaxies (these are selected
according to their offset from the model/observed main sequence). In
particular, we focus on the differences between central and satellite galaxies
with the aim to determine when and how these differences arise.

The overall distributions of HI and H$_2$ masses, SFRs, SF and stellar
  radii of model galaxies agree relatively well with those of observed
  galaxies.  Comparing the median scaling relations of central and satellite
  star forming galaxies separately, we find that the different data-sets
  considered are consistent with each other. The model reproduces reasonably well
  the measured HI mass-stellar mass relation, stellar size-stellar mass
  relations for both central and satellite galaxies, while predicting a lower
  normalization for the H$_2$ mass-stellar mass, SFR-stellar mass, and SF
  size-stellar mass relations.

For the HI mass, H$_2$ mas, SFR, and SF radii, the measured differences between
central and satellite galaxies are $\sim 0.2$, $\sim 0.5$, $\sim 0.5$, $\sim
0.1$, respectively. No significant difference is measured for the stellar
radii. The model agrees well with the observational data for the differences 
in HI mass, and SF/stellar radii, while over-predicting significantly the
differences in H$_2$ mass and SFR.  For our model galaxies, we use the
available galaxy merger trees to verify if differences between central and
satellite galaxies result from environmental processes or originate before
environment starts playing a role. We find that all differences considered
can be ascribed to environmental effects, which reduces to stripping of the
hot-gas reservoir in our model.

The stellar and gaseous sizes of satellite galaxies in our model are comparable
to those of their central counterparts at both accretion time and present
time. This is due to the similar assembly history of their host haloes, that is a result of the selection/matching adopted for central-satellite pairs. In
our model, the size growth of star forming-galaxies is dominated by cooling in
the case of the gaseous stellar discs, and by subsequent star formation for the
stellar discs. Mergers and disc instabilities play a minor role in the size 
growth of our model galaxies. After accretion (i.e. the time when a central 
galaxy is accreted onto a larger halo, becoming a satellite galaxy), sizes
stop growing because of the suspension of cooling and of the low fraction of stars
formed. Meanwhile, central galaxies grows very little at late time.  

Including only strangulation, our model reproduces well the median 
observed HI masses, SF radii, and stellar radii for both central and satellite 
main sequence (star forming) galaxies.
In contrast, it tends to over-predict the depletion of molecular gas and the
related suppression of the star formation activity. We argue that this could
be largely resolved with the inclusion of a proper treatment for ram-pressure
stripping of the cold gas and for non-instantaneous stripping of hot gas. A treatment of angular momentum balance that
accounts for the multi-phase nature of the gaseous disc is also required. We
plan to work on these aspects in the future.

\section*{Acknowledgements}
LX and GDL acknowledge financial support from the MERAC foundation.  DW and MF
acknowledge the support of the Deutsche Forschungsgemeinschaft (DFG) under
Project ID 3871/1-1 and ID 3871/1-2. We thank Alessandro Boselli for the help 
on using the HRS data. We also thank the anonymous referee for comments and 
suggestions that improve the manuscript significantly.




\bibliographystyle{mnras}
\bibliography{size}


\appendix




\bsp	
\label{lastpage}

\end{document}